\documentclass[a4paper,aps,prb,twocolumn,amsmath,amssymb,showpacs,10pt]{revtex4}
\usepackage{epsfig}
\usepackage{graphicx}
\usepackage{amsmath,amssymb}

\newcommand{\be}{\begin{equation}}
\newcommand{\ee}{\end{equation}}

\newcommand{\Neel}{\text{N}{\acute{\textrm{e}}}\text{el}}

\begin{document}
\title{Structural phase transitions in geometrically frustrated antiferromagnets}
\author{T. E. Saunders and J. T. Chalker}
\affiliation{Theoretical Physics, University
  of Oxford, 1 Keble Road, Oxford, OX1 3NP, United Kingdom}
\date{\today}
\begin{abstract}
We study geometrically frustrated antiferromagnets with magnetoelastic
coupling. Frustration in these systems may be relieved by a structural
transition to a low temperature phase with reduced lattice
symmetry. We examine the statistical mechanics of this transition 
and the effects on it of quenched disorder,  
using Monte Carlo simulations of the classical Heisenberg model on the
pyrochlore lattice with coupling to uniform lattice distortions.
The model has a transition between a cubic, paramagnetic
high-temperature phase and a tetragonal, N\'eel ordered low-temperature
phase. It does not support the spin-Peierls phase, 
which is predicted as an additional possibility within Landau theory,
and the transition is first-order for reasons unconnected with the
symmetry analysis of Landau theory. Quenched
disorder stabilises the cubic phase, and we find a phase diagram as a
function of temperature and disorder strength similar to that
observed in Zn$_{1-x}$Cd$_x$Cr$_2$O$_4$.
\end{abstract}


\maketitle

\section{Introduction}
\label{Intro}
Much of the interesting behaviour observed in geometrically frustrated
antiferromagnets \cite{reviews} can be related to macroscopic ground state degeneracy
in classical models for these systems. Some of the simplest models,
with only nearest neighbour exchange interactions, remain in the
paramagnetic phase down to zero
temperature.\cite{villain,reimers,moessner1998} Characterising the 
strength of exchange interactions by the magnitude of the Curie-Weiss
constant $\Theta_{\rm CW}$, the paramagnetic phase is strongly
correlated at temperatures $T\ll |\Theta_{\rm CW}|$. More realistically,
ground state degeneracy is lifted by additional interactions or coupling between
spins and other degrees of freedom. As a result, the magnet may order
or freeze at a temperature $T_{\rm c}$. If the relevant energy scales
are small compared to exchange, $T_{\rm c} \ll |\Theta_{\rm CW}|$. One
such escape route from the strongly correlated paramagnet arises from
magnetoelastic coupling, which may lead to a magnetically driven
cooperative Jahn-Teller
transition.\cite{lee2000,Yamashita2000,Tchernyshyov2002a,Tchernyshyov2002}
An alternative is 
that weak quenched disorder leads to spin freezing at low
temperature.\cite{Saunders2007} In this paper we study structural phase transitions
induced by magnetoelastic coupling in geometrically frustrated
antiferromagnets and their suppression by quenched disorder.

Recent theoretical work
\cite{Yamashita2000,Tchernyshyov2002a,Tchernyshyov2002} on magnetic
Jahn-Teller transitions in 
geometrically frustrated magnets has combined an exact treatment of a
tetrahedral cluster of four spins with Landau theory applied to the
pyrochlore lattice. Results for the cluster show that arbitrarily weak
magnetoelastic coupling generates an instability to a distorted ground
state. Landau theory elevates this instability to an ordering
transition, with several possible phase
diagrams.\cite{Tchernyshyov2002a,Tchernyshyov2002} Depending on the 
distortion mode, the transition from the cubic, paramagnetic phase to
a tetragonal phase may be a continuous one, or it may be first order.
If the transition is continuous, then spin Peierls order is expected in the
tetragonal phase immediately below the critical point, but
may give way to N\'eel order at lower temperature. Alternatively, if the
cubic to tetragonal transition is discontinuous, it may either be to 
a phase with spin Peierls order, or directly to one with
N\'eel order. 

While Landau theory provides an exhaustive
listing of possibilities, a microscopic approach is necessary to
decide which occur in a given model. Moreover, it may happen that transitions which
on symmetry grounds need not be first order, because there is no cubic
invariant in the Landau expansion for the free energy, in fact are
discontinuous, because a fourth-order invariant has a negative
coefficient. These considerations motivate the work we describe
here. 

We examine as a representative model the classical Heisenberg
antiferromagnet with nearest neighbour exchange interactions on the pyrochlore
lattice, including magnetoelastic coupling to distortions from cubic
to tetragonal symmetry. Ordering occurs if a lattice distortion is
accompanied by a free energy gain in the spin system that outweighs
its cost in elastic energy. The ordering temperature $T_{\rm c}$
depends on the strength of the magnetoelastic coupling compared to the
appropriate elastic 
modulus. If coupling is weak, $T_{\rm c} \ll |\Theta_{\rm CW}|$.
In the weak-coupling regime, thermodynamic behaviour is independent of
the strength of exchange interactions and a universal
function of the ratio $T/T_{\rm c}$. We investigate the balance 
between magnetic free energy and elastic energy in this regime by using
Monte Carlo simulations of the spin system to provide a numerically exact
treatment of its statistical mechanics. We combine this with a mean
field treatment of the lattice degrees of freedom, in which we allow
only uniform lattice distortions. We believe our approach brings the
advantages of a microscopic treatment of the magnetic degrees of
freedom that drive ordering, without the complications and additional
parameters that would arise in a full treatment of phonons.

Our results establish a rather different picture of the magnetic
Jahn-Teller transition from that suggested previously. We find that the
structural transition is strongly first order, and to a N\'eel ordered
state, independently of any cubic invariant in the elastic energy. In this sense, the transition is
driven by the energy gain arising from N\'eel order, and not from spin
Peierls order. Since the jump we observe in the N\'eel order parameter
is large, we believe this conclusion is likely to be robust,
even though a change in microscopic details of the model may in principle
favour spin Peierls order.

The fact that arbitrarily weak magnetoelastic coupling is sufficient
to generate a ground state instability prompts one to ask why
structural phase transitions are not ubiquitous in geometrically
frustrated antiferromagnets. We suggest here that weak exchange disorder
(which is generated by random strains in the presence of magnetoelastic coupling)
acts to stabilise the high symmetry lattice structure. As we have
shown elsewhere, in the absence of a structural transition, there is
spin glass ordering.\cite{Saunders2007} In this sense, the magnetic Jahn-Teller and spin
glass transitions are alternative low-temperature fates for a
geometrically frustrated antiferromagnet.

The remainder of the paper is organised as follows. In
Sec.~\ref{model} we define our model and
summarise our main conclusions. We present a Monte Carlo study of the
thermally-driven transition in a clean system in Sec.~\ref{thermal}, and
consider the disorder-driven transition at zero temperature in
Sec.~\ref{disorder}. Finally, in Sec.~\ref{discussion} we discuss our
results in relation to experiments on 
Zn$_{1-x}$Cd$_x$Cr$_2$O$_4$.

\section{Model and summary of results}
\label{model}

The model we treat has degrees of freedom of two types: classical Heisenberg
spins ${\bf S}_i$ of unit magnitude at the sites $i$ of a pyrochlore
lattice; and a uniform tetragonal distortion of this lattice,
parameterised by a single variable $\Delta$. We include four 
contributions to the energy of the system: the exchange energy ${\cal
  H}_0$ of the undistorted lattice; the change in exchange energy ${\cal
  H}_{\Delta}$ due to magnetoelastic coupling; the elastic energy ${\cal H}_{\rm el}$; and the
energy ${\cal  H}_{\rm dis}$ arising from quenched
exchange disorder. 

With interaction strength $J$ 
the exchange energy of the cubic pyrochlore lattice is
\begin{equation}
{\cal H}_0 = J \sum_{ij} {\bf S}_i \cdot{\bf S}_j\,,
\end{equation} 
where the sum is over nearest neighbour pairs. This Hamiltonian by
itself is known to have a macroscopically degenerate ground state and
to remain in the paramagnetic phase down to $T=0$.\cite{reimers,moessner1998} Our concern here is
to find how other interactions lift this degeneracy and induce low
temperature phase transitions.

\begin{figure}[tbh]
\includegraphics[width=8cm]{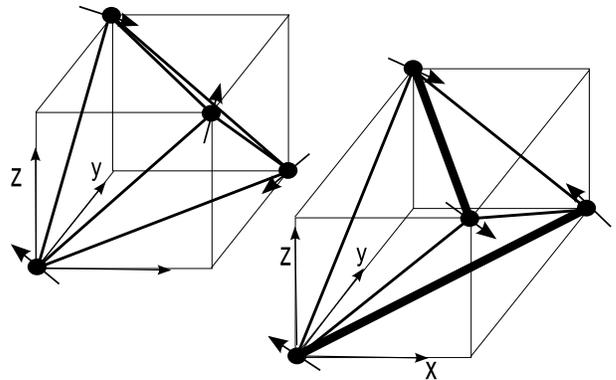}
\caption{\label{fig:model} Classical groundstate spin configurations
on a single tetrahedron. 
Left:  for an undistorted tetrahedron all exchange interactions are
of equal strength and ground states include non-collinear
spin configurations as illustrated. 
Right: for a tetragonally distorted tetrahedron, two longer
bonds (broad lines) 
and four shorter bonds result in modulated interaction strengths, and
ground state spin configurations are collinear.}
\end{figure}

We express the change in exchange energy due to magnetoelastic coupling
in terms of the variable $\Delta$. 
For definiteness, we take the lattice distortion to be a compression of the cubic $z$ axis
and an expansion of the $x$ and $y$ axes. We characterise the
amplitude of this distortion by its effect on exchange interactions,
taking the interaction strength in the distorted system to be
$J-\Delta$ between neighbouring spin pairs in the same $x$-$y$
plane, and $J+\Delta/2$ between pairs in adjacent $x$-$y$
planes. Then the magnetoelastic coupling is
\begin{equation}
{\cal H}_{\Delta} = \sum_{ij} \Delta_{ij} {\bf S}_i \cdot{\bf S}_j\,,
\end{equation}
where $\Delta_{ij}$ has the values $-\Delta$ and $\Delta/2$ for
pairs $i,j$ of the two types. The ground states of ${\cal H}_0 + {\cal
  H}_{\Delta}$ for $\Delta >0$ are N\'eel ordered: all spins in each
$x$-$y$ plane of the lattice are ferromagnetically aligned, with
opposite orientations on successive planes, as illustrated in
Fig.~\ref{fig:model}. (The opposite choice of sign, $\Delta <0$,
leads to classical ground states with high degeneracy, and we do not
consider it further.)   

We include elastic energy within a harmonic approximation, writing for a system of $N$ spins
\begin{equation}
{\cal H}_{\rm el} = \frac{\kappa}{2} N \Delta^2\,.
\label{magnetoelastic}
\end{equation}
The coupling constant $\kappa$ is related to an elastic coefficient
$c$ and the derivative $\partial J/\partial x$ of exchange
interaction strength with distance in the following way. First, 
$\Delta$ can be expressed in terms of a strain $\epsilon$ and an
equilibrium bond length $a$, as the product $\Delta = \epsilon a\, \partial
J/\partial x$. Taking the elastic energy density to be $c
\epsilon^2/2$, we obtain Eq.~(\ref{magnetoelastic}) with $\kappa \sim
ca/(\partial J/\partial x)^2$. Thus $\kappa$ is large if
magnetoelastic coupling is weak.

Exchange disorder is represented by
\begin{equation}
{\cal H}_{\rm dis} = \sum_{ij} \delta_{ij} {\bf S}_i \cdot {\bf S}_j
\end{equation}
with $\delta_{ij}$ non-zero only for nearest neighbours, and drawn
independently for each such pair $i,j$ from a distribution that is
uniform over the range 
$[-\delta,\delta]$. 

\begin{figure}[tbh]
\includegraphics[width=8cm]{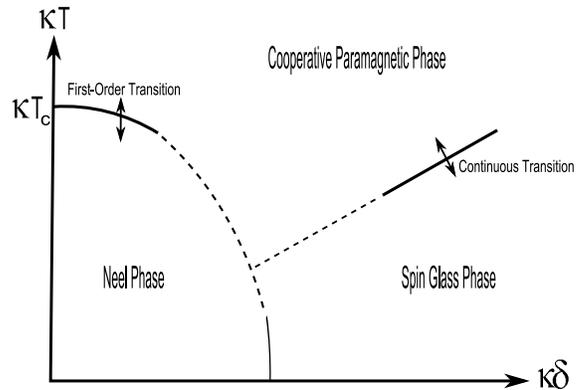}
\caption{\label{fig:pd} Schematic phase diagram in the plane of scaled
  temperature $\kappa T$ and disorder strength $\kappa
  \delta$, based on numerical simulations of
  (\ref{Hamiltonian}). $T_c$ denotes the temperature of 
  a transition between a $\Neel$ ordered, tetragonal low temperature
  phase, and a paramagnetic, high temperature phase in the pure
  system. A similar transition occurs at zero temperature, at a
  disorder strength $\delta_{\rm c}$. Solid lines represent phase
  boundaries  that have been probed in simulations. Dashed lines are
  guides to the eye.}  
\end{figure}

Our objective is to study the model defined by the Hamiltonian
\begin{equation}
{\cal H} = {\cal H}_0 + {\cal H}_{\Delta} + {\cal H}_{\rm el} +
{\cal H}_{\rm dis}
\label{Hamiltonian}
\end{equation}
in the regime $\kappa \gg J^{-1}$ and $\Delta,\,\delta \ll J$. In these circumstances
${\cal H}_0$ makes the dominant contribution to the energy of the
system and transitions from the paramagnetic phase take place at
temperatures $T \ll J$ (we set $k_{\rm B}=1$ throughout this paper).
In Fig.~\ref{fig:pd} we show a schematic phase diagram for the model 
as a function of the dimensionless variables $\kappa T$ and
$\kappa \delta$, which we now discuss.
First, consider the system without quenched disorder ($\delta = 0$) and
at $T=0$. Ground states are N\'eel ordered for $\Delta>0$ and in
these states ${\cal H}_{\Delta} = -2N\Delta$. As a result, $\cal H$ is
minimised by a non-zero tetragonal lattice distortion, with $\Delta =
2/\kappa$, no matter how weak the magnetoelastic coupling and hence
how large the coefficient $\kappa$. The net energy gain per spin from this lattice distortion is
$2/\kappa$, and so one expects a transition to the paramagnetic phase
at a temperature $T_{\rm c} \sim \kappa^{-1}$. As we demonstrate in Sec.~\ref{thermal}, this
transition is first order. Second, consider disruption of N\'eel order
by exchange randomness at $T=0$. Strong disorder leads to a larger
modulation of exchange interactions than that produced by a uniform
lattice distortion. In turn, this results in ground states with spin
configurations that are random rather than N\'eel ordered, and an
energy that is minimised at $\Delta=0$. The transition from the N\'eel
ordered state takes place at
$\delta \sim \kappa^{-1}$. We show in Sec.~\ref{disorder} that it is probably
continuous. Finally, consider the effect of non-zero temperature at
strong disorder. We have studied the model without magnetoelastic
coupling ($\kappa \to \infty$) elsewhere, finding a phase transition
from a spin-glass ordered low temperature phase to a paramagnetic high
temperature phase, with a transition temperature $T_{\rm g} \sim
\delta$.\cite{Saunders2007} We include this phase boundary in Fig.~\ref{fig:pd}, although
we have not attempted to investigate the multicritical point at which
the three phases meet.

\section{Magnetically-driven structural phase transition}
\label{thermal}

In this section we discuss the thermally driven transition in a system
without disorder, between a high temperature, cubic, paramagnetic
phase and a low-temperature, tetragonal phase, which we will show is
N\'eel ordered. We
restrict our attention to the regime of weak magnetoelastic coupling,
$\kappa \gg J^{-1}$, in which behaviour depends only on the variable
$\kappa T$ and is independent of $J$. Our main tool is the use of
Monte Carlo simulations to determine the free energy of the spin
system.

Let $F(\Delta,T)$ be the free energy per spin at temperature $T$,
calculated from the Hamiltonian ${\cal H}_0 + {\cal H}_{\Delta}$. 
The combination
\be
\label{G0}
\Gamma(\Delta, T) = \frac{{\cal H}_{\rm el}}{N} + F(\Delta,T) - F(0,T) 
\ee
is the net free energy gain per spin arising from a lattice distortion.
The physical value of $\Delta$ is the one that minimises
$\Gamma(\Delta,T)$, and the system is in a tetragonal phase if there is
a value of $\Delta>0$ for which $\Gamma(\Delta,T)<0$.

The contribution of the spin degrees of freedom to this free energy
for $T,\Delta \ll J$ can be written in the scaling form
\begin{equation}
F(\Delta,T) - F(0,T) = \Delta f_{\rm s}(\Delta/T)\,, 
\end{equation}
since in this regime $\Delta$ sets the
only energy scale for the spin system.
The function $f_{\rm s}(x)$ plays an important role in the following,
and it is useful to discuss its low and high temperature
asymptotics. At low temperature ($x \gg 1$) the magnetic free energy
difference is dominated by the energy gain $2\Delta$ per spin in the N\'eel state, 
and so $\lim_{x \to \infty} f_{\rm s}(x) = -2$. Conversely, a
high-temperature expansion (covering the regime $1\ll \kappa T \ll
\kappa J$) gives for small $x$
\begin{eqnarray}
N\Delta f_{\rm s}(\Delta/T) = \langle {\cal H}_{\Delta} \rangle_0
&-&\frac{1}{2T}[ \langle ({\cal H}_{\Delta})^2 \rangle_0 - \langle
  {\cal H}_{\Delta} \rangle_0^2] \nonumber\\ &+& {\cal O}(\Delta^3/T^2) \,,
\end{eqnarray}
where $\langle \ldots \rangle_0$ denotes an average with the
Hamiltonian ${\cal H}_0$ in the limit $T\ll J$. This average yields
$\langle {\cal H}_{\Delta} \rangle_0=0$ and $\langle {\cal H}_{\Delta}
\rangle_0^2 =bN\Delta^2$ with $b>0$, and so $f_{\rm s}(x) \sim -bx/2$ for
$x\to 0$.  

Now Eq.~(\ref{G0}) can be written in terms of $f_{\rm s}(x)$ as 
\begin{eqnarray}
\Gamma(\Delta, T) &=& \frac{\kappa}{2} \Delta^2 + \Delta f_{\rm s}(\Delta/T) \nonumber\\
	&=& T \left[ \frac{\kappa T}{2} x^2 + xf_{\rm s}(x) \right] \,.
 \label{G1}
\end{eqnarray}
The form of the function $f_{\rm s}(x)$ determines
the critical point $\kappa T_{\rm c}$ and the order of the transition.
The equation $\kappa T x/2+ f_{\rm s}(x) =0$ has no solution for positive $x$ in the
cubic phase ($\kappa T > \kappa T_{\rm c}$) but has a solution $x_{\rm
  c}>0$ in the tetragonal phase. If $x_{\rm c}$ approaches zero as
$\kappa T$ approaches $\kappa T_{\rm c}$ from below, the transition is
continuous, while if $x_{\rm c}$ remains finite in this limit, the 
transition is first order. Thus, the transition is continuous if $f_{\rm s}(x)$ is convex for
$0<x<\infty$; otherwise it is first order.

\subsection{Numerical Approach}

To find the magnetic free energy difference between the cubic and
tetragonal phases, and specifically the function $f_{\rm s}(x)$, we use Monte
Carlo simulations of a spin system with the Hamiltonian ${\cal H}_0 +
{\cal H}_{\Delta}$. The system simulated had the shape of a
rhombohedron, with edges parallel to the primitive
basis vectors of the lattice and with
periodic boundary conditions.  We focus on the temperature range $T\ll J$ for fixed
$\Delta \ll J$ and ensure equilibration at low temperature by
employing parallel tempering.\cite{parallel} 
Representative simulation parameters are $\Delta/J = 10^{-3}$ and $T/J
\geq 10^{-4}$, with system sizes $32 \leq N \leq 2048$ and measurement run
lengths of up to $2\times 10^5$ parallel tempering steps, preceded by
equilibration for the same number of steps. Within the parallel
tempering algorithm spin configurations are exchanged amongst a range of
possible temperatures: we checked in particular that each
configuration explored all temperatures across an interval that spanned the first-order
transition temperature. The data presented here are predominantly for systems of $500$ and $864$
spins, but larger system sizes were also simulated to ensure that finite size
effects are not significant. 

Determination of the free energy requires calculation of the entropy
$S(T)$. We do this by integration of the heat capacity $C_{\rm v}(T)$,
using 
\be
\label{entropy}
S(T_0) - S(T) = \int^{T_0}_{T} \frac{C_{\rm v}(T)}{T}\,{\rm d}T
\ee
and choosing the reference temperature $T_0 \gg \Delta$ so that
$S(T_0)$ is independent of $\Delta$ and
cancels from the free energy difference $F(\Delta,T) - F(0,T)$.
We note that since
$C_{\rm v}(T)$ is constant at low temperature in our classical model,
$S(T)$ is logarithmically divergent as $T\to 0$, but the contribution
$-TS(T)$ to the free energy has no divergence.
For fixed $\Delta > 0$ the spin system has a N\'eel ordering
transition, which is first order in the regime of interest, $\Delta
\ll J$. Numerical integration of $C_{\rm v}(T)/T$ across this
transition is difficult, because the latent heat of the transition
appears in a finite system as a narrow spike in $C_{\rm v}(T)$. 
We circumvent this difficulty by using two independent approaches to
find the entropy change at the transition, and checking between
them for consistency. One approach is direct integration, using spline
fits of $C_{\rm v}(T)$ around the transition. The other is to
use the change in internal energy at the transition to determine the
latent heat and hence the entropy jump.

\subsection{Results}

\begin{figure}[tbh]
\includegraphics[width=8cm]{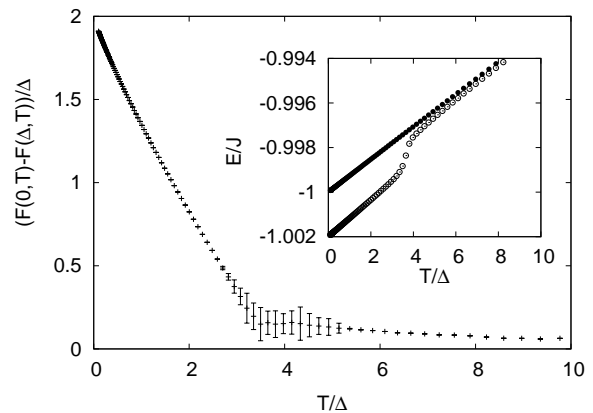}
\caption{\label{fig:free} Scaled free energy difference
  $(F(0)-F(\Delta))/\Delta$ vs $T/\Delta$ for $\Delta/J =
  10^{-3}$. Inset: internal energy per spin on the same temperature
  scale, for $\Delta/J=10^{-3}$ ($\circ$) and $\Delta=0$
  ($\bullet$). All data for $N=256$ spins.}
\end{figure}

Results from simulations for the scaled free energy difference
$[F(\Delta,T)-F(0,T)]/\Delta$ and internal energy $E$ per spin as a
function of temperature are shown in Fig.~\ref{fig:free}. We discuss
first the behaviour of $E$. For a system with $\Delta=0$ the ground
state energy per spin is $-J$, and $E$ increases linearly with $T$ at low
temperature, because $C_{\rm v}(T)$ is constant. For $\Delta>0$ the
ground state energy per spin is reduced by $2\Delta$. For fixed
$\Delta$ there is a N\'eel transition, which takes place at $T/\Delta
\approx 3.1$: at the transition the internal energy per spin increases
abruptly and approaches its value in a system with
$\Delta=0$. Consider next the the free energy difference. It is also small 
in the high temperature phase. In contrast to the internal energy
difference, the free energy difference is of course continuous through
the transition. It grows with decreasing temperature
below the transition.

\begin{figure}[tbh]
\includegraphics[width=8cm]{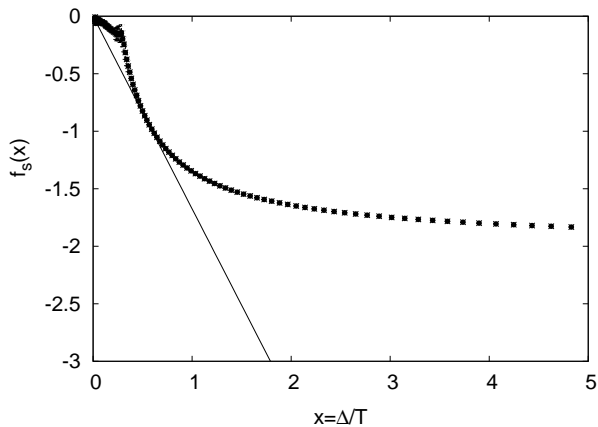}
\caption{\label{fig:fs} $f_{\rm s}(x)$ vs $x=\Delta/T$ for
  fixed $\Delta/J=10^{-3}$ ($\bullet$). The straight line is tangent to
  $f_{\rm s}(x)$ and has gradient $-1.67$. } 
\end{figure}

The free energy difference is replotted as a function of the variable
$x=\Delta/T$ in Fig.~\ref{fig:fs}, together with the straight line
through the origin that is tangent to the function $f_{\rm s}(x)$. 
From the gradient of this line we identify the transition temperature
$T_{\rm c}$ in the system with magnetoelastic coupling, obtaining
$\kappa T_{\rm c} = 3.34$. We also see that the transition is strongly
first order, since the solution $x_{\rm c}\approx 0.60$ to the equation 
$f_{\rm s}(x) = - \kappa T_{\rm c} x/2$ is finite. Note that the
transition for the full model with Hamiltonian $\cal H$, in which the
value of $\Delta$ varies with $T$ to minimise the free energy,
preempts the one in a system with fixed $\Delta$. The latter
transition is responsible for the kink in $f_{\rm s}(x)$, visible at
$x \approx 0.3$ in Fig.~\ref{fig:fs}. Of course, it is the extra
elastic energy accompanying a lattice distortion that destabilises the
N\'eel ordered state at a lower temperature in the full model than in
the one with $\Delta$ fixed.

\begin{figure}[tbh]
\includegraphics[width=8cm]{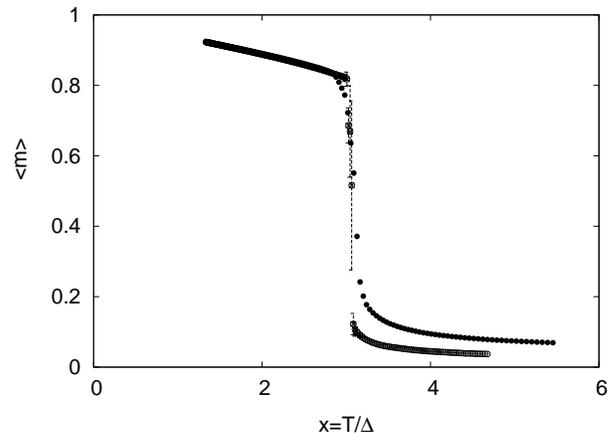}
\caption{\label{fig:neel} Sublattice magnetisation
  $m$ vs $T/\Delta$ with $\Delta/J= 10^{-3}$ for $N=864$ ($\bullet$)
  and 4000 ($\circ$).  Errors only shown for $N=4000$ for clarity.} 
\end{figure}

To demonstrate that the tetragonal phase is indeed N\'eel ordered, we
show in Fig.~\ref{fig:neel} the variation of the sublattice
magnetisation with temperature in a system with a fixed value of
$\Delta$. At the critical point of the full model, $T/\Delta \approx
1.7$ and the sublattice magnetisation is approximately 90\% of its
saturation value. We discuss the transition at fixed $\Delta$ in
detail elsewhere.\cite{pickles}

In summary, the simple approach we have presented, which combines a
microscopic treatment of the statistical mechanics of the spin
degrees of freedom with mean field theory for the lattice distortion, 
yields a strongly first-order transition between cubic and tetragonal
phases, with highly developed N\'eel order at all temperatures within
the tetragonal phase. A first-order transition is also predicted
using Landau theory, but the mechanism in that case is different: it
arises because symmetry permits a term in the elastic energy which is
cubic in strain. In our approach such a term would appear as a
contribution to ${\cal H}_{\rm el}$ cubic in $\Delta$. 
Within Landau theory, first order transitions can arise in the absence
of cubic invariants, if the expansion of the free energy in powers of
the order parameter has a quartic term with a negative
coefficient. This appears to be the mechanism that leads to the first
order transition we find.

\section{Influence of quenched disorder}
\label{disorder}

As we have seen, a structural phase transition is a generic feature of
the model we have studied, irrespective of the strength of
magnetoelastic coupling. The driving mechanism is the linear gain in
exchange energy with lattice distortion for some ground state spin
configurations of the high-symmetry structure. This magnetic Jahn-Teller mechanism is
likely to apply to a variety of other classical models for geometrically frustrated
magnets. Specifically, for it to be effective the set of ground
states should include configurations in which ${\bf S}_i \cdot {\bf
  S}_j$ takes different values for different nearest neighbour spin pairs,
so that the exchange energy of the state varies linearly with lattice
distortion $\Delta$. For example, this condition includes Heisenberg
model on the lattice appearing in the material SCGO,\cite{ramirez1990} but excludes
the kagome Heisenberg antiferromagnet, in which ${\bf S}_i \cdot {\bf
  S}_j = -1/2$ for all nearest neighbours in all ground states. In
many materials, however, no structural transition is
observed. Instead, geometrically frustrated magnets often show spin 
freezing at low temperature. Such freezing suggests the importance of
residual disorder, even though samples in some cases appear to be almost
disorder-free. In this section we examine how disorder may stabilise
the cubic phase of our model. We present results obtained in two
ways. First, we use Monte Carlo simulations to study a spin system with
fixed $\Delta > 0$ at various disorder strengths $\delta$ as a
function of temperature. We find that strong disorder suppresses the
N\'eel transition that is observed at $\delta = 0$. Second, we examine
behaviour at $T=0$ as a function of disorder strength, treating the
dependence of ground state energy on $\delta$ in the same way as we
did the dependence of free energy on $T$ in Sec.~\ref{thermal}.

\subsection{Behaviour for $T>0$ at fixed $\Delta$}

The introduction of random exchange interactions complicates Monte Carlo
simulations. Longer equilibration times are required (up to $10^6$
parallel tempering steps) and we must average over disorder
configurations. This limits the system sizes that can be simulated.  
The results we present were obtained for systems of size $N=864$,
averaging over 100 realisations for each disorder strength, with
$\Delta/J = 10^{-2}$. 

\begin{figure}[tb]
\includegraphics[width=8cm]{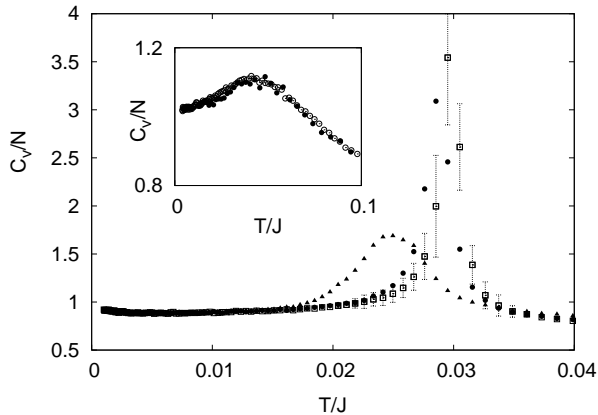}
\caption{\label{fig:cv} Heat capacity per spin vs $T/J$ for
  $\Delta/\delta=10(\square), 1 (\bullet), \textrm{ and } 0.5
  (\blacktriangle)$ in a system of 864 spins at
  $\Delta/J=10^{-2}$. Errors shown for system with largest error. Inset: heat capacity per spin vs $T/J$
  for $500$ ($\bullet$) and $864$ ($\circ$) spins at
  $\Delta/\delta=10^{-1}$ and $\Delta/J=10^{-2}$. Errors indicated by scatter of points. } 
\end{figure}

We show in Fig.~\ref{fig:cv} the heat capacity
per spin as a function of temperature for various disorder strengths,
expressed in terms of $\Delta/\delta$ since we expect this ratio to
control behaviour in the regime $\Delta,\delta,T\ll J$. We find that
the sharp peak in heat capacity, associated at $\delta = 0$ with a
discontinuous N\'eel ordering transition, occurs at lower temperature
with increasing disorder but persists for $\Delta/\delta
\geq 1$. At larger disorder strength, $\Delta/\delta <1$, it is
rounded, being smooth and independent of system size at $\Delta/\delta
= 10^{-1}$ (see inset to Fig.~\ref{fig:cv}). The dominant errors are from disorder
averaging. We conclude that the N\'eel transition is suppressed by
disorder. It is presumably replaced at strong disorder by a spin glass
transition, but because of the difficulties attached to Monte Carlo
simulations in this regime, we do not investigate this further.

\subsection{Behaviour at $T=0$}
\label{zero-temp}

To make additional progress in quantifying the effect of disorder on the
structural and magnetic phase transitions, we focus on behaviour at
$T{=}0$ as a function of $\delta$. Let $E(\Delta,\delta)$ be the
disorder-averaged ground state energy per spin for the spin system
with Hamiltonian ${\cal H}_0 + {\cal H}_\Delta +{\cal H}_{\rm dis}$. In
analogy with Eq.~(\ref{G0}), we define the energy combination
\begin{equation}
\Gamma_{\rm E}(\Delta,\delta) = \frac{{\cal H}_{\rm el}}{N} + E(\Delta,\delta) -
E(0,\delta)\,.
\end{equation}
To determine the ground state lattice symmetry for a given disorder
strength $\delta$, the value of $\Delta$ should be chosen to minimise
$\Gamma_{\rm E}(\Delta,\delta)$. In the cubic phase the minimum is at
$\Delta=0$, while in the tetragonal phase it is at $\Delta >0$. 

We expect the spin contribution to this energy difference in the limit
$\Delta,\delta \ll J$ to have the scaling form
\begin{equation}
E(\Delta,\delta) - E(0,\delta) = \Delta f_{\rm E}(\Delta/\delta)
\label{eqn:Delta-E}
\end{equation}
with the asymptotic behaviour for $f_{\rm E}(y)$ given at large $y$ by $\lim_{y\to\infty}
f_{\rm E}(y) = -2$, and at small $y$ by $f_{\rm E}(y) \propto y$. Using
this scaling form
\begin{equation}
\Gamma_{\rm E}(\Delta,\delta) = \Delta\left[\frac{\kappa \delta}{2} y +
f_{\rm E}(y)\right]\,.
\end{equation}
Thus the state of the system is controlled by
the dimensionless coupling constant $\kappa\delta$. As for the
thermally driven transition without disorder, there is a critical
coupling $\kappa \delta_{\rm c}$. If $\kappa \delta > \kappa
\delta_{\rm c}$ the equation $\kappa \delta y/2 + f_{\rm E}(y) =0$ has
no solution for $y>0$ and the system is in the cubic
phase. Alternatively, if  $\kappa \delta < \kappa \delta_{\rm c}$ 
there is a solution at $y_{\rm c} > 0$ and the system is in the
tetragonal phase. As for the thermally driven transition, the
disorder-driven one is continuous if $f_{\rm E}(x)$ is convex for
$0<x<\infty$ and first order otherwise.

\begin{figure}[tb]
\includegraphics[width=8cm]{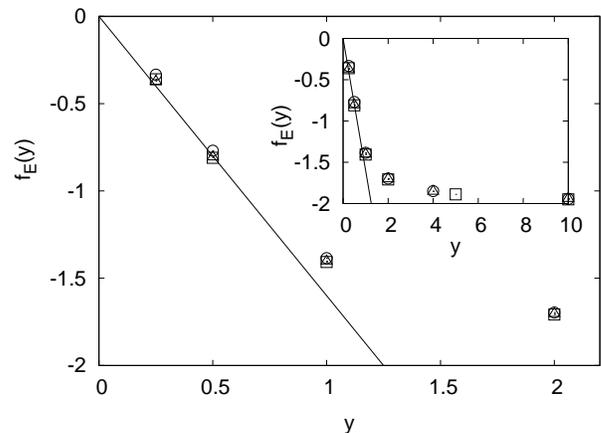}
\caption{\label{fig:fe} $f_E(y)$ vs $y$ for three values of
  $\Delta/J$: $10^{-1}$ ($\circ$), $10^{-2}$ ($\diamond$) and
  $5\times10^{-3}$ ($\square$). The straight line is tangent to $f_{\rm
    E}(y)$ at $y=0$ and has gradient $-\kappa \delta_c/2 = - 1.6$ The
  inset shows the same plot but for a larger range of $y$. Errors are
  less than the symbol sizes.} 
\end{figure}

We show in Fig.~\ref{fig:fe} the function $f_{\rm E}(y)$ determined
numerically for a system of size $N=864$ spins, averaging over 50
disorder realisations. Low energy configurations are obtained by
quenching low temperature states taken from Monte Carlo simulations
using parallel tempering. The agreement between data for three values of
$\Delta/J$ demonstrates that the calculations are within the scaling
regime $\Delta,\delta \ll J$. In contrast to behaviour at the
thermally driven transition, it appears from the form of $f_{\rm
  E}(y)$ that $y_{\rm c}$ will approach zero smoothly as $\kappa
\delta$ approaches $\kappa \delta_{\rm c}$ from below. Hence the
disorder driven, zero temperature transition is probably continuous. We find 
$\kappa \delta_{\rm c} = 3.2$.

\begin{figure}[tb]
\includegraphics[width=8cm]{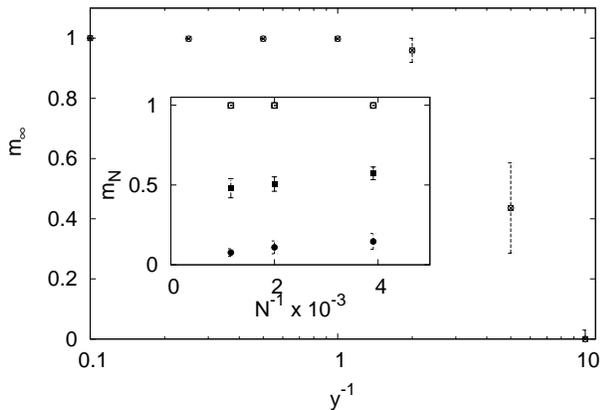}
\caption{\label{fig:op} Staggered magnetisation per spin in the
  infinite system vs $y^{-1}$ on a logarithmic scale. Inset:
  illustration of the extrapolation of sublattice magnetisation $m$ vs
  $N^{-1}$ to infinite system size.} 
\end{figure}

As a supplement to this study of the structural transition, we also
examine the variation of sublattice magnetisation $m$ with disorder
strength at fixed $\Delta$ in ground states. The sublattice
magnetisation shows significant finite-size effects in our range of
system sizes, and so we extrapolate to the limiting value $m_{\infty}$ 
from results for $N=256$, $500$ and $864$. We show
the variation of $m_{\infty}$ with $y$ in Fig.~\ref{fig:op}, and the
extrapolation in the inset. Sublattice magnetisation is not much
affected by disorder for $y>1$, and decreases to zero for $y\approx
0.17$. If it is indeed the case that $f_{\rm E}(y)$ is convex, so that
the structural transition is continuous, and if it is also the case
that the magnetisation is zero below a finite value of $y$, then the
disorder driven, zero temperature transition is a two-stage one, and
there exists a tetragonal phase without N\'eel order at intermediate
disorder strength. A more detailed investigation of this point would
be demanding, and we do not pursue it.

\subsection{Paramagnetic-Spin Glass Phase Transition}
\label{spinglass}

We have not investigated the regime $\kappa \delta \gg 1$ as part of
the present work, but we draw attention to earlier
studies of the effect of weak disorder ($\delta \ll J$) at $\Delta=0$
(in effect, $\kappa\to\infty$).
\cite{bellier-castella2001,Saunders2007} 
A continuous phase transition between the 
paramagnetic and spin glass phases occurs at a temperature
proportional to $\delta$.\cite{Saunders2007}

\section{Discussion}
\label{discussion}

The model we have discussed provides a caricature for the chromium
spinel oxides ACr$_2$O$_4$, where A is Mg,\cite{rovers} Zn,\cite{lee2000} Cd,
\cite{rovers} or Hg.\cite{ueda2006} These $S=3/2$ 
pyrochlore antiferromagnets undergo structural phase transitions
accompanied by N\'eel ordering, with transition temperatures that are
small compared to the magnitude of their Curie-Weiss constants. The
structural distortions and magnetic ordering patterns are 
different in each compound and are considerably
more complex than the one we have considered.\cite{ueda2006,lee2007} While we are therefore
not able to make a detailed comparison with experiment, some
qualitative points deserve emphasis. These concern the mechanism
driving the transition, the order of the transition, and the influence
of non-magnetic impurities.

Our calculations demonstrate that, at least in the classical limit and
for the uniform distortion mode we consider, the mechanism driving the
structural transition is the energy gain from N\'eel order, rather
than spin Peierls order. It is possible in principle that other
distortion modes would allow a phase at intermediate temperature that has
spin Peierls but no N\'eel order, but the very large discontinuity we
observe in the N\'eel order parameter suggests to us that this is
unlikely. It is also possible that quantum fluctuations favour spin
Peierls order, and it would be interesting to attempt in future
research to quantify this. 

In addition, our results show that the phase transition can be
first order for a reason unconnected to the symmetry analysis of
Landau theory. This is significant because within Landau theory some
structural distortion patterns are expected to lead to a first order
transition while others should not. In particular, within Landau
theory a transition involving a tetragonal distortion that is
predominantly an odd mode (affecting the two tetrahedra in a unit cell
oppositely) should be continuous, while a transition involving an even
mode should be first order.\cite{Tchernyshyov2002a,Tchernyshyov2002}
Our simulations demonstrate that a 
first order transition may result from a microscopic treatment,
independently of expectations based on symmetry.

Finally, it is interesting to compare our study of the influence of
exchange randomness on the transition with observations of the effect
of non-magnetic impurities in the material
Zn$_{1-x}$Cd$_x$Cr$_2$O$_4$.\cite{Ratcliff2002} In this system a surprisingly small 
level of substitution ($x=0.03$) at non-magnetic sites is sufficient
to suppress N\'eel order and the structural phase transition. The
origin of this behaviour is believed to be the larger ionic radius of
Cd$^{2+}$ compared to Zn$^{2+}$: as a consequence, substitution leads
to random strains, which in the presence of magnetoelastic coupling
generate exchange randomness. The results we present in
Sec.~\ref{disorder} show this mechanism in action, and are
qualitatively consistent with the phase diagram observed in
Zn$_{1-x}$Cd$_x$Cr$_2$O$_4$ as a function of temperature and disorder
strength, with $x$ playing the role of our variable $\delta$.\cite{Ratcliff2002}

We thank P. C. W. Holdsworth, R. Moessner, and  O. Tchernyshyov for
valuable discussions. The work was supported in part by EPSRC Grant
No. GR/R83712/01.

\end{document}